
\input phyzzx
\hsize 420pt
\hoffset 24.8775pt
\def\gt{\tilde g}
\def\Rt{\tilde R}

\def\pr#1#2#3#4{Phys. Rev. D {\bf#1}, #2 (19#3#4)}
\def\prl#1#2#3#4{Phys. Rev. Lett. {\bf#1}, #2 (19#3#4)}
\def\pl#1#2#3#4{Phys. Lett. {\bf#1}, #2 (19#3#4)}

\def\sqr#1#2{{\vcenter{\hrule height.#2pt
        \hbox{\vrule width.#2pt height#1pt \kern#1pt
                \vrule width.#2pt}
        \hrule height.#2pt}}}

\def\Sa{S_{\alpha}}
\def\ro{\rho_0}
\def\ri{\rho_1}
\def\fo{\phi_0}
\def\phii{\phi_1}
\def\rbar{{\bar\rho}}
\def\roo{{{\bar\rho}_0}}
\def\rd{{\Delta{\bar\rho}}}

\nopagenumbers
\line{\hfill CU--TP--661}
\line{\hfill hep-ph/9411410}
\vskip 1in
\centerline{\fourteenbf Bubble Nucleation Rate In Fourth-Order Gravity}
\vskip .6in
\settabs 3\columns
\+&Yue Hu \hfill&\cr
\+&Department of Physics\hfill&\cr
\+&Columbia University\hfill&\cr
\+&New York, NY 10027\hfill&\cr
\vfill
\midinsert
\narrower
\centerline{Abstract}
We calculate the bubble nucleation rate in a fourth-order gravity theory
whose action contains an $R^2$ term, under the thin-wall approximation,
by two different methods.
First the bounce solution is found for the
Euclidean version of the original action.
Next, we use a conformal transformation to transform the
theory into general relativity with an additional scalar field
and then proceed to find the bounce solution. The same results
are obtained from both calculations.

\endinsert

\eject

\pagenumbers
\pagenumber=1

\chapter{ Introduction }

Consider the theory of a single scalar field defined by the action
$$  S=\int d^4x\left[{1\over2}\partial_\mu\phi\/\partial^\mu\phi-V(\phi)\right]
		\eqn\action  $$
where $V$ has two local minima, $\phi_\pm$. One of these, $\phi_-$,
is a global minimum (``true vacuum''). The other corresponds to a
metastable state (``false vacuum'')
that decays through bubble nucleation.
The decay rate from the false vacuum to the true vacuum
per unit volume can be calculated in the
semiclassical limit
$$ \Gamma/V=Ae^{-B/\hbar}\bigl[1+O(\hbar)\bigr]. \eqno\eq $$
Algorithms for computing the coefficients $B$ and $A$ have been
given by Coleman[1] and Callan and Coleman[2] respectively.

In applications to the early universe, the effect of gravity
becomes interesting.
When gravity is weak, the basic features of bubble nucleation are similar
to those in flat space, with small modification of the decay rate
and the bubble radius.
If we confine ourselves to the case where
both the true and false vacua have positive energy density, then the
true and false vacua will be de Sitter spaces. Gravitation makes the
materialization of the bubble more likely, and makes the radius of the
bubble at its moment of materialization smaller. Coleman and
De~Luccia[3] have calculated the coefficient $B$ in the case where
gravity is described by general relativity.

The Hilbert action for
general relativity contains a single term proportional to
the curvature scalar, and leads to second-order field equations.
For some modified theories of gravity, the action contains
terms that are combinations
of the curvature scalar, Ricci tensor, Weyl tensor and the full
Riemann curvature tensor. This leads to higher-order field equations.
Such theories have been considered (see, e.g., [7]), for example,
in the context of variations of extended inflation [8, 9].
In this paper, we will confine ourselves
to the case where the gravitational action contains an arbitrary function
of the curvature scalar. The method can be easily applied to
the case where the action contains arbitrary functions of the curvature
scalar multiplied by a function of another scalar field.

When there is only an additional $R^2$ term in the action, Whitt found
a conformal transformation that transforms the theory to
standard general relativity but introduces a new scalar
field[5].
This method can be generalized to the case of an arbitrary function
of the curvature scalar.
This is done in Section II, where some general features
of fourth-order gravity are discussed.
To calculate the bubble nucleation rate in fourth-order gravity theory,
one could use the Euclidean version of the original action.
This action contains higher powers of the curvature scalar. The bounce
solution[1] of this Euclidean action gives the coefficient $B$ of
the bubble nucleation rate. Alternatively
one could use a conformal transformation to transform the theory
into general relativity with an additional scalar field. One then could
proceed following Coleman and De~Luccia[3].
In Section III, we calculate the coefficient $B$ when
the extra terms are small, using a perturbation method.
In Section IV, we will use the conformal transformation to calculate the
coefficient $B$. The result of this approach is found to agree
exactly with the direct approach.
We have made no attempt to calculate the coefficient $A$. Even in general
relativity, there is not a satisfactory method to evaluate this
coefficient.

\vfill\eject

\chapter{ Fourth-Order Gravity and Conformal Transformation }

Consider a theory described by the action
$$  S=\int d^4x\sqrt{-g}\left(-{f(R)\over16\pi G}+{\cal L}_{\rm m}
     \right)  \eqn\actf  $$
where $R$ is the curvature scalar and
${\cal L}_{\rm m}$ is the Lagrangian of the matter fields. General
relativity is the case when $f(R)=R$. For simplicity,
we will confine ourselves to only one matter field with the matter Lagrangian
$$ {\cal L}_{\rm m}={1\over2}\partial_\mu\phi\ \partial^\mu\phi-V(\phi)
   \eqno\eq.  $$
The generalization to more fields is straightforward.
It is easy to show that the equations of motion for this theory are
$$\eqalign{ & {1\over2}f(R)g_{\mu\nu}-f'(R)R_{\mu\nu}+f''(R)
  (R_{;\mu\nu}-{R_{;\alpha}}^\alpha g_{\mu\nu}) \cr
 & +f'''(R)(R_{;\mu}R_{;\nu}-R_{;\alpha}R^{;\alpha}g_{\mu\nu})
   +8\pi G(\partial_\mu\phi\ \partial_\nu\phi-{\cal L}_{\rm m}g_{\mu\nu})=0  }
  \eqn\mot $$
and
$$ {\phi_{;\lambda}}^{;\lambda}=-{dV(\phi)\over d\phi}.  \eqno\eq  $$
Because the curvature contains second order derivatives of the metric,
the above equations contain fourth-order derivatives of the metric
unless $f''(R)\equiv0$ or, in other words, if the theory is general relativity.
However, the higher order terms enter the equation only through the
derivatives of the curvature scalar. We can lower the order of the
differential equations by two if we introduce a new variable $\sigma$
and arrange for this to be equal to the curvature scalar.
The equations are then second order, but we have to supply the equation
$$ \sigma(x)=R(g_{\mu\nu}) \eqno\eq  $$
for the new variable.

This can also be done
by looking at the action directly. The action~\actf\ is equivalent to
$$ S=\int d^4x\sqrt{-g}\left\{-{1\over16\pi G}\bigl[f'(\sigma)R
      -\sigma f'(\sigma)+f(\sigma)\bigr]+{\cal L}_{\rm m}\right\}. \eqn\acts $$
By varying $\sigma$ we get
$$ \sigma=R  \eqn\same.  $$
Substituting this relation back into the action~\acts, we find that
it is exactly
the same as the original action~\actf. To make the theory look like
general relativity, we introduce the conformal transformation
$$ \gt_{\mu\nu}=f'(\sigma)g_{\mu\nu}. \eqn\factor $$
In the new metric the connection is
$$ {\tilde \Gamma}^\lambda_{\mu\nu}=\Gamma^\lambda_{\mu\nu}
  +\Sigma^\lambda_{\mu\nu}  \eqno\eq  $$
where the tensor $\Sigma^\lambda_{\mu\nu}$ is
$$  \Sigma^\lambda_{\mu\nu}={f''(\sigma)\over2f'(\sigma)}
   \left[\sigma_{,\mu}g^\lambda_\nu+\sigma_{,\nu}g^\lambda_\mu
     -\sigma^{,\lambda}g_{\mu\nu}\right].  \eqno\eq  $$
The Ricci tensor in the new metric is found to be
$$\eqalign{ \Rt_{\mu\nu}= & R_{\mu\nu}-{f''(\sigma)\over2f'(\sigma)}\left[
  2\sigma_{;\mu\nu}+{\sigma^{;\lambda}}_{;\lambda}g_{\mu\nu}\right] \cr
 & -{f'''\over2f'}\left[2\sigma_{,\mu}\sigma_{,\nu}+\sigma_{,\lambda}
   \sigma^{,\lambda}g_{\mu\nu}\right]+{3\over2}\left({f''\over f'}\right)
   \sigma_{,\mu}\sigma_{,\nu}  } \eqno\eq  $$
and the new curvature scalar is
$$\eqalign{ \Rt= & \Rt_{\mu\nu}\gt^{\mu\nu} \cr
 = & {1\over f'(\sigma)}\left\{R-{3f''\over f'}{\sigma^{,\lambda}}_\lambda
   -{3f'''\over f'}\sigma_{,\lambda}\sigma^{,\lambda}+{3\over2}\left(
   {f''\over f'}\right)^2\sigma_{,\lambda}\sigma^{,\lambda}\right\}  }
   \eqn\cur  $$
where the indices are raised and lowered by the old metric.
Also,
$$\eqalign{ {\sigma^{;\lambda}}_\lambda|_{\rm old}=\sigma_{;\mu;\nu}g^{\mu\nu}
  &=(f'{\sigma^{;\lambda}}_\lambda-f''\sigma_{,\lambda}\sigma^{,\lambda})|_
  {\rm new}   \cr
  \sigma_{,\lambda}\sigma^{,\lambda}|_{\rm old} & = f'
  \sigma_{,\lambda}\sigma^{,\lambda}|_{\rm new} }   \eqno\eq   $$
With the aid of the above relation, the action can be written as
$$\eqalign{ S=\int d^4x\sqrt{-\gt} &  \left\{-{\Rt\over16\pi G}+{1\over16\pi G}
  \left[{3\over2}\left({f''\over f'}\right)^2\sigma_{,\lambda}\sigma^{,\lambda}
  +{\sigma\over f'}-{f\over{f'}^2}\right]\right. \cr
  & \ \ \ \  \left. +{\partial_\mu\phi\ \partial_\nu\phi\ \gt^{\mu\nu}\over2f'}
    -{V(\phi)\over{f'}^2} \right\}  }  \eqno\eq   $$
where the arguments of $f$, $f'$ and $f''$ are $\sigma$. Now we have the action
of general relativity coupled to
two scalar fields, although the matter Lagrangian
is somewhat unconventional. It can be shown that the equations of
motion for the
new action are equivalent to those of the original action, which proves the
legitimacy of the above procedure.

Before we go to the next section, let us look at the de Sitter
space solution of fourth-order gravity. De Sitter space is a
homogeneous space with $O(4,1)$ symmetry.
Because of this symmetry, the Ricci tensor
can be written as
$$ R_{\mu\nu}={1\over4}Rg_{\mu\nu} \eqno\eq  $$
where $R$ is a constant throughout the whole space-time.
The matter field is also homogeneous throughout the
space-time manifold, so the equation of motion~\mot\ reduces to
$$ {1\over4}f'(R)R-{1\over2}f(R)=\kappa V(\phi_0)  \eqn\const $$
where $\phi_0$ is a stationary point of $V$ and
$\kappa=8\pi G$. This is an algebraic equation
which can be solved for $R$.

Let us consider the specific case where the action is
$$  S=\int d^4x\sqrt{-g}\left\{-{R\over16\pi G}+\alpha R^2+{1\over2}
    \partial_\mu\phi\ \partial^{\mu}\phi-V(\phi)\right\}. \eqn\acta $$
Then $f(\sigma)$ is
$$  f(\sigma)=\sigma-16\pi G\alpha\sigma^2,  \eqno\eq  $$
and the conformal transformation is
$$ \gt_{\mu\nu}=(1-4\alpha\kappa\sigma)g_{\mu\nu}  \eqn\cnfmfctr $$
In this new metric the action is
$$   S=\int d^4x\sqrt{-\gt}\left\{-{\Rt\over2\kappa}
 +{12\alpha^2\kappa\partial_\mu\sigma\partial^\mu\sigma
     \over(1-4\alpha\kappa\sigma)^2}
  +{\partial_\mu\phi\ \partial^\mu\phi\over2(1-4\alpha\kappa\sigma)}
   -{V(\phi)+\alpha\sigma^2\over
       (1-4\alpha\kappa\sigma)^2}\right\}.   \eqn\actc  $$

For both general relativity and the theory described by~\acta,
equation~\const\ gives
$$ R=-4\kappa V(\phi_0)   \eqno\eq  $$
because the $\alpha$ terms happen to cancel.
If higher order terms enter the action, the
result differs in general from the general relativistic result.
In the case when only $R$ and $R^2$ terms enter,
this solution corresponds to the stationary point of the
last term of~\actc,
$$	U(\sigma)={V(\phi_0)+\alpha\sigma^2\over(1-4\alpha\kappa\sigma)^2}
		\eqno\eq $$
This stationary point is located at
$\sigma=-4\kappa V(\phi_0)$, which is just the curvature
scalar in the original metric. The curvature scalar in the new metric is
$$  \Rt=-{4\kappa V(\phi_0)\over1+16\alpha\kappa^2 V(\phi_0)} \eqno\eq  $$
in agreement with Eq.~\cur.

\vfill\eject

\chapter{Bounce Action: The Direct Approach}

Consider a theory described by the action~\actf, where $V(\phi)$ has two
local minima, $\phi_\pm$, only one of which, $\phi_-$, is an absolute minimum.
Let us also assume that both $V(\phi_-)$ and $V(\phi_+)$ are positive.
The classical field theory defined by this action possesses two stable
homogeneous equilibrium states, $\phi=\phi_+$ and $\phi=\phi_-$.
In the quantum
version of the theory, though, only the second one corresponds to a truly
stable state, a true vacuum. The first decays through barrier penetration; it
is a false vacuum. The decay rate per unit time per unit volume
can be calculated in the semiclassical approximation as
$$ \Gamma/V=Ae^{-B/\hbar}(1+O(\hbar)).   \eqno\eq  $$

To calculate the coefficient $B$, let us consider the Euclidean version of the
theory. The Euclidean action is defined as minus the formal analytic
continuation to imaginary time of the Lorentzian action~\acta\
$$ S_E=\int d^4x\sqrt{g}\left\{-{f_E(R)\over16\pi G}+{1\over2}\partial_\mu
   \phi\ \partial_\nu\phi\ g^{\mu\nu}+V(\phi)\right\}  \eqn\actb  $$
where the metric is the usual positive-definite one of Euclidean
four-space. Let $(\phi, g)$ be a solution of the Euler-Lagrange equations
associated with $S_E$ such
that: [i] $(\phi, g)$ approaches the false vacuum solution at large Euclidean
distance, [ii] $(\phi, g)$ is not identical to the false vacuum solution, and
[iii] $(\phi, g)$ has Euclidean action less than or equal to that of any other
solution obeying [i] and [ii]. Then the coefficient $B$ in the vacuum decay
amplitude is given by
$$  B=S_E(\phi, g)-S_E(\phi_+, g_+)  \eqno\eq $$
where $(\phi_+,g_+)$ is the false vacuum solution. $(\phi, g)$ is called the
bounce solution.

It can be shown that in flat space the bounce is always $O(4)$-symmetric[4]. In
curved space, although not proven, it is very plausible that the bounce is
still $O(4)$-symmetric. We will work under this assumption.
The metric can then be written as
$$ ds^2=d\xi^2+\rho^2(\xi)d\Omega^2  \eqn\coor  $$
where $\xi$ is a radial coordinate from the center of the $O(4)$-symmetry and
$\rho(\xi)$ measures the circumference divided by $2\pi$ at radial coordinate
$\xi$.

In the following, we will confine ourselves to the case where only $R$ and
$R^2$ terms enter the action and the theory is described by the action~\acta.
In $O(4)$-symmetric coordinates~\coor, the equations of motion are
$$\eqalign{  {\rho'}^2=1+{\kappa\rho^2\over3}&\left\{{1\over2}{\phi'}^2
              -V \right.  \cr
   &+\left. 36\alpha\left[{1+2{\rho'}^2-3{\rho'}^4\over\rho^4}
          +{2{\rho'}^2\rho''\over\rho^3}-{{\rho''}^2\over\rho^2}
           +{2\rho'\rho'''\over\rho^2}\right]\right\} } \eqn\motrho  $$
and
$$ \phi''+{3\rho'\over\rho}\phi'={dV\over d\phi}. \eqn\motphi $$

We will use perturbative methods to find the effect of small $\alpha$.
To be self-consistent, the second term in the bracket should be much
smaller than the first term. This is always true if
$$ \alpha{m^2\over M_p^2}\ll1$$
because
the second term is at most of the order $\alpha m^6/M_p^2$, where $m$ is the
mass scale of the potential $V(\phi)$.

Under this condition, the Euclidean action can be split into an unperturbed
part
$$ S_0=\int d^4x\sqrt{g}\left\{-{R\over2\kappa}+{1\over2}(\partial\phi)^2
   +V(\phi)\right\}   \eqno\eq  $$
and a perturbation
$$ \Sa=\int d^4x\sqrt{g}(-\alpha R^2).  \eqno\eq  $$

Let $(\ro,\fo)$ be the bounce solution when $\alpha$ vanishes and
$(\ri,\phii)$ be a small correction
$$\eqalign{ \rho&=\ro+\ri \cr \phi&=\fo+\phii \cr} \eqno\eq $$
Because $(\ro,\fo)$ is a solution to the
Euler-Lagrange equations of $S_0$, the first order correction to $S_0$ due
to $(\ri,\phii)$ vanishes, so to first order the Euclidean action is
$$  S_E=S_0[\ro,\fo]+\Sa[\ro,\fo]+O(\alpha^2)  \eqno\eq  $$
and the bounce action is
$$\eqalign{  B&=B_0+B_\alpha \cr
	&=\bigl\{S_0(\ro,\fo)-S_0({\rm false\ vacuum})\bigr\}
	+\bigl\{\Sa(\ro,\fo)-\Sa({\rm false\ vacuum})\bigr\} } \eqno\eq $$

We will use the thin wall approximation in the following calculation[1,3].
The thin wall approximation is valid when the energy density difference between
the false and true vacua is small compared to the potential barrier. In this
case the thickness of the wall is small compared to the radius of the bubble.
We will also assume that gravity is weak, i.e., that the scale $m$ of $V(\phi)$
is much smaller than the Planck mass. The Euclidean true and false vacua with
positive cosmological terms are Euclidean de~Sitter spaces, which are
four-spheres. We will further assume that the radii of these four-spheres are
much larger than the flat-space bubble radius.
Under above assumptions the bounce solution has a thin 3-dimensional spherical
wall which separates an interior true vacuum region and an exterior false
vacuum region. The radius of this wall is large compared to the wall thickness,
but much smaller than the radii of the true and false vacuum 4-spheres.

First let us consider the unperturbed solution with $\alpha=0$. Inside
(outside)
the bubble $(\ro,\fo)$ is just de Sitter space of true (false) vacuum
$$\eqalign{ \fo&=\phi_\pm\cr \ro&={\sin(H_\pm\xi)\over H_\pm} } \eqn\unsol $$
with
$$  R_\pm=12H_\pm^2=4\kappa V(\phi_\pm). \eqn\unr $$
This is the exact result for the true and false vacuum 4-spheres even when
$\alpha\neq0$ because the $\alpha$ terms happen to cancel.
The contributions to $B_0$ have been calculated by Coleman and De Luccia[3]
$$\eqalign{ \left.B_0\right|_{\rm out} &=0 \cr
	\left. B_0\right|_{\rm in} &= {12\pi^2\over\kappa^2V(\phi_-)}
 \left\{\left[1-{1\over3}\kappa\roo^2V(\phi_-)\right]^{3/2}-1\right\}
  -(\phi_-\rightarrow\phi_+) } \eqno\eq $$
where $2\pi\roo$ is the circumference of the 3-sphere thin wall.
In the thin wall approximation, the contribution to $B_0$ from the shell
can be shown to be[3]
$$ \left.B_0\right|_{\rm wall}=4\pi^2\roo^3\int d\xi\bigl[V(\phi)-V(\phi_+)
	\bigr] =2\pi^2\roo^3S_1 \eqno\eq $$
where $\roo$ is determined by requiring that the bounce action
$$ B_0=\left.B_0\right|_{\rm in}+\left.B_0\right|_{\rm wall}
	+\left.B_0\right|_{\rm out}   $$
be stationary.

We need the curvature scalar in the wall region for latter calculations.
Because
$(\ro, \fo)$ is a solution to the theory defined by $S_0$ (which is standard
Euclidean general relativity), $(\ro,\fo)$ satisfies Einstein's equation
$$ R_{\mu\nu}-{1\over2}g_{\mu\nu}R=-\kappa T_{\mu\nu} \eqno\eq $$
The trace of this,
$$ R=\kappa {T^\lambda}_\lambda, \eqno\eq  $$
gives
$$  R=\kappa\left[{\fo'}^2+4V(\phi_0)\right].  \eqno\eq  $$
Because the difference between $V(\phi_+)$ and $V(\phi_-)$ is very small
compared to $V(\phi)$ in the thin-wall region, define
$$\eqalign{ V_0 &={1\over2}\bigl[V(\phi_+)+V(\phi_-)\bigr] \cr
	U(\phi)&=V(\phi)-V_0 } \eqno\eq $$
Except in the thin wall region, $\phi$ is almost
a constant (either $\phi_+$ or $\phi_-$) and $V$ is well approximated by $V_0$.
Because the bubble's circumference radius is large,
we can neglect the ${3\rho'\over\rho}\phi'$ term in
Eq.~\motphi. We then have the ``approximate conservation law''
$$  {1\over2}{\fo'}^2-V(\fo)=-V_0.  \eqno\eq  $$
We find
$$\eqalign{ \fo' & =\sqrt{2U(\fo)}  \cr
             R & = \kappa\bigl[6U(\fo)+4V_0\bigr]. } \eqn\wallcur  $$

Now consider the first order correction to the coefficient $B$
$$  B_\alpha=S_\alpha(\ro,\fo)-S_\alpha({\rm false\ vacuum}) \eqno\eq $$
In the exterior of the bubble
the bounce solution coincides with the false vacuum solution, so
the contribution to $B_\alpha$ vanishes
$$  \left.B_\alpha\right|_{\rm out}=0.  \eqno\eq  $$
The contribution of the wall region to $B_\alpha$ is
$$  \left.B_\alpha\right|_{\rm wall}=-\alpha\int d^4x\left(R^2-R^2_{\rm false}
	\right)  \eqno\eq  $$
Using \wallcur\ we find that
$$\eqalign{ \left.B_\alpha\right|_{\rm wall}&=(-\alpha\kappa)2\pi^2\roo^3
 \int d\xi\ 6U\bigl(6U+8V_0\bigr) \cr
 &=(-\alpha\kappa)2\pi^2\roo^3\int{d\phi\over\sqrt{2U}}
	\ 6U\bigl(6U+8V_0\bigr)\cr
  &= -12\alpha\pi^2\kappa^2\roo^3
  \int_{\phi_-}^{\phi_+}d\phi\ \sqrt{2U}(3U+4V_0). }  \eqn\bwall  $$

The contribution from the interior of the bounce to $B_\alpha$ is
$$\eqalign{\left.B_\alpha\right|_{\rm in}&=\int d^4x(-\alpha R^2)_{\rm true}
 -\int d^4x(-\alpha R^2)_{\rm false} \cr
 &=-2\pi^2\alpha\left\{\int d\xi\rho^3(4\kappa V)^2\Big|_{\rm true}
  -\int d\xi\rho^3(4\kappa V)^2\Big|_{\rm false}\right\} }  $$
Using equations \unsol and \unr\ we find
$$\eqalign{\left.B_\alpha\right|_{\rm in}&=-32\alpha\pi^2\kappa^2\left\{
	\int d\xi\ {\sin^3(H_-\xi)\over H_-^3}V_-^2
	-\int d\xi\ {\sin^3(H_+\xi)\over H_+^3}V_+^2\right\} \cr
	&= 96\alpha\pi^2   \left[
  \sqrt{1-{\kappa V_-\over3}{\roo}^2}\left(2+{\kappa V_-\over3}
  {\roo}^2\right)
   -\sqrt{1-{\kappa V_+\over3}{\roo}^2}\left(2+{\kappa V_+\over3}
   {\roo}^2 \right) \right] \cr}     \eqn\bin $$
where $V_-$ and $V_+$ are the energy densities of the true and false vacua
respectively.

Combining the contributions from the three regions, we find that to first order
in $\alpha$
$$\eqalign{ B= & B_0 -12\alpha\pi^2\kappa^2{\roo}^3
  \int_{\phi_-}^{\phi_+}d\phi\sqrt{2U}(3U+4V_0) \cr
   &  +96\alpha\pi^2   \left[
  \sqrt{1-{\kappa V_-\over3}{\roo}^2}\left(2+{\kappa V_-\over3}
  {\roo}^2\right)
   -\sqrt{1-{\kappa V_+\over3}{\roo}^2}\left(2+{\kappa V_+\over3}
  {\roo}^2 \right) \right].  }   \eqn\bounce $$

Now we want to determine the sign of the first order correction to the bounce
action. If it is positive, the bubble nucleation will be less likely. First
notice that
$$ \int_{\phi_-}^{\phi_+}d\phi\ 3U\sqrt{2U}>0 \eqno\eq  $$
If the cosmological term $V_0$ is positive, then $B_\alpha|_{\rm wall}$~\bwall
\ has the opposite sign from $\alpha$ and
$$  \Bigl|\left.B_\alpha\right|_{\rm wall}\Bigr|>
	48|\alpha|\pi^2\kappa^2\roo^3V_0\int_{\phi_-}^{\phi_+}d\phi\ \sqrt{2U}
		\eqno\eq  $$
Under our assumption of weak gravity, $\kappa V\roo^2$ is small and we expand
$B_\alpha|_{\rm in}$ to the leading order in it
$$ \left.B_\alpha\right|_{\rm in}
	=8\alpha\pi^2\kappa^2\roo^4V_0(V_+-V_-) \eqno\eq $$
This has the same sign as $\alpha$ if $V_0$ is positive. Therefore the
contributions to the first order correction from the wall region and from
the bubble interior have opposite signs. To compare their magnitudes
$$\left|{\left.B_\alpha\right|_{\rm wall}\over\left.B_\alpha\right|_{\rm in}}
  \right|>{6\int d\phi\sqrt{2U}\over\roo(V_+-V_-)}  \eqno\eq  $$
notice that to leading order (see [1,3])
$$ \roo={3\over(V_+-V_-)}\int d\phi\ \sqrt{2U}={3S_1\over(V_+-V_-)} \eqno\eq $$
So we conclude
$$ \eqalign{ &
\left|{\left.B_\alpha\right|_{\rm wall}\over\left.B_\alpha\right|_{\rm in}}
   \right|>2  \cr
  & {B_\alpha\over\alpha}<0 \cr } \eqn\b $$
Similarly, for
$$ {B_\alpha}'={dB_\alpha(\roo)\over d\roo} \eqno\eq  $$
we have
$$ \eqalign{ &
\left|{\left.{B_\alpha}'\right|_{\rm wall}
	\over\left.{B_\alpha}'\right|_{\rm in}}\right|>{3\over2} \cr
  & {{B_\alpha}'\over\alpha}<0 \cr} \eqn\bd $$

The bubble radius
$\rbar$ is determined by requiring that $B$ be stationary at that point.
Let $\roo$ be the unperturbed bubble radius
and $\rd$ the first order correction
$$ \rbar=\roo+\rd+O(\alpha^2). \eqno\eq  $$
We require that
$$ 0=\left.{\partial B\over\partial\rho}\right|_\rbar
  =\left.{\partial B_0\over\partial\rho}\right|_\rbar+\left.
  {\partial B_\alpha\over\partial\rho}\right|_\rbar   \eqn\bal  $$
To first order in $\alpha$, we have
$$ \left.{\partial B_0\over\partial\rho}\right|_\rbar
 \approx\left.{\partial B_0\over\partial\rho}\right|_{\roo}+ \left.{\partial^2
B_0\over\partial\rho^2}\right|_{\roo}\rd
 =\left.{\partial^2B_0\over\partial\rho^2}\right|_{\roo}\rd   \eqno\eq  $$
and
$$ \left.{\partial B_\alpha\over\partial\rho}\right|_\rbar
 \approx\left.{\partial B_\alpha\over\partial\rho}\right|_{\roo} \eqno\eq $$
Then equation \bal\ requires
$$ \rd=-\left.{{B_\alpha}'\over{B_0}''}\right|_{\roo} \eqno\eq $$
As mentioned above, ${B_\alpha}'$ is negative when $\alpha$ is positive. From
the work by Coleman and De Luccia[3] and the work by Parke[6], we know that
${B_0}''$ is negative. Therefore when $\alpha$ is positive, $\rd$ is negative,
which means that bubble has a smaller circumference radius than it does when
the $R^2$ term is absent in the Lagrangian.

The above conclusion can be understood intuitively. Because of the gradient
term in the Lagrangian, the inhomogeneity in the wall region creates
surface tension and increases the bubble energy.
In order to be energetically balanced, the bubble interior
should have a lower potential energy density to compensate the energy increase
due to surface tension.
For small $\alpha$, the
theory can be treated as general relativity with the effective potential
$$ V_{\rm eff}=V-\alpha R^2  \eqno\eq $$
The presence of the $R^2$ term lowers the potential. Both the
true and false vacua
are de Sitter spaces and the curvature scalar is proportional to the energy
density. For the false vacuum the energy density is larger than for the
true vacuum, so the relative energy difference is reduced by the presence of
the $R^2$ term.
However in the thin wall region the curvature scalar becomes
fairly big when the field changes rapidly. This means that in the wall region
the potential energy decreases at a much larger rate than the potential energy
inside the bubble. So both the surface tension and the interior potential
energy difference decrease, but the surface tension decreases at a larger rate.
So to be balanced energetically, the volume inside the bubble should decrease.
This is why the bubble radius decreases. Generally, it is easier for a small
bubble to nucleate, so the bounce action decreases and the bubble nucleation
rate increases.

\vfill\eject

\chapter{Bounce Action: Conformal Transformation Method}

In this section, we are going to use the conformal transformation of
Section~II to study the bubble nucleation rate in fourth-order gravity.
If the Lorentzian action is Eq.~\acta, then the Euclidean action is
$$   S_E=\int d^4x\sqrt{g}\left\{-{R\over16\pi G}-\alpha R^2+
  {1\over2}(\partial\phi)^2+V(\phi)\right\}.  \eqno\eq  $$
The conformal transformation described in Section~II leads to the action
$$ S_E=\int d^4x\sqrt{\gt}\left\{-{m_p^2\over16\pi}\Rt+
  {12\alpha^2\kappa\partial_\mu\sigma\partial^\mu\sigma\over
  (1+4\alpha\kappa\sigma)^2}
 +{\partial_\mu\phi\ \partial^\mu\phi\over2(1+4\alpha\kappa\sigma)}
  +{V(\phi)+\alpha\sigma^2\over
     (1+4\alpha\kappa\sigma)^2}\right\}.  \eqno\eq  $$
Assuming $O(4)$ symmetry, we use the metric~\coor. The
Euclidean action then becomes
$$\eqalign{ S_E=2\pi^2\int d\xi & \left\{\rho^3\left[{{\phi'}^2\over
   2(1+4\alpha\kappa\sigma)}
 +{12\alpha^2\kappa{\sigma'}^2\over(1+4\alpha\kappa\sigma)^2}
 +{V+\alpha\sigma^2\over(1+4\alpha\kappa\sigma)^2}\right]\right.  \cr
 + & \left.{3\over\kappa}(\rho^2\rho''+\rho{\rho'}^2-\rho)\right\}. }
    \eqno\eq  $$
The Einstein equation is
$$ {\rho'}^2=1+{1\over3}\kappa\rho^2\left[{{\phi'}^2\over
    2(1+4\alpha\kappa\sigma)}
   +{12\alpha^2\kappa{\sigma'}^2\over(1+4\alpha\kappa\sigma)^2}
   -{\alpha\sigma^2+V\over(1+4\alpha\kappa\sigma)^2}\right].  \eqn\mmotrho $$
The other equations of motion are
$$ \phi''+{3\rho'\over\rho}\phi'={{dV\over d\phi}+4\alpha\kappa\sigma'\phi'
   \over 1+4\alpha\kappa\sigma}   \eqn\mmotphi  $$
and
$$ \sigma''+{3\rho'\over\rho}\sigma'={8\alpha\kappa{\sigma'}^2\over1+
     4\alpha\kappa\sigma}
 -{{\phi'}^2\over6\alpha}+{1\over6\alpha\kappa}{\sigma-4\kappa V\over
   1+4\alpha\kappa\sigma}.  \eqn\mmotsigma  $$
When $\alpha=0$, these equations of motion agree with Eqs.~\motrho\ and
\motphi.

The second derivative of $\rho$ can be eliminated by integration by parts.
(The surface term from the parts integration is harmless because we are only
interested in the action difference between two solutions that agree at the
boundary.) We thus obtain
$$\eqalign{  S_E=2\pi^2\int d\xi & \left\{\rho^3\left[{{\phi'}^2\over2(1+
   4\alpha\kappa\sigma)}
 +{12\alpha^2\kappa{\sigma'}^2\over(1+4\alpha\kappa\sigma)^2}
 +{V+\alpha\sigma^2\over(1+4\alpha\kappa\sigma)^2}\right]\right. \cr
   & \left. -{3\over\kappa}
     (\rho{\rho'}^2+\rho)\right\}. } \eqno\eq  $$
We now use Eq.~\mmotrho\ to eliminate $\rho'$. We find
$$ S_E=4\pi^2\int d\xi\left[-{3\over\kappa}\rho+\rho^3
  {V+\alpha\sigma^2\over(1+4\alpha\kappa\sigma)^2}\right]. \eqn\bb  $$

Now we evaluate the above integral in three regions. Outside the bubble the
bounce solution agrees with the false vacuum and the
contribution to the bounce action is zero:
$$   B_{\rm out}=0   \eqno\eq   $$
The interior of the bubble is de Sitter space and from the equations of
motion we know
$$\eqalign{ \sigma &= 4\kappa V  \cr
     \rho & ={\sin(H\xi)\over H} } \eqno\eq  $$
with
$$  H^2={\kappa V\over3(1+16\alpha\kappa^2V)}.  \eqno\eq  $$
The contribution to the bounce action is the difference of the integral~\bb\
between the true vacuum and false vacuum. It is easy to find that
$$ B_{\rm in}={12\pi^2\over\kappa^2V}(1+16\alpha\kappa^2V)\left\{
  \left[1-(H\rbar)^2\right]^{3/2}-1\right\}_{true}-(false)  \eqno\eq  $$
Expanding this to first order in $\alpha$, we find
$$\eqalign{ B_{\rm in}= &B_{\rm in}^0(\rbar) \cr
	& +96\alpha\pi^2   \left[
  \sqrt{1-{\kappa V_-\over3}{\rbar}^2}\left(2+{\kappa V_-\over3}
  {\rbar}^2\right)
   -\sqrt{1-{\kappa V_+\over3}{\rbar}^2}\left(2+{\kappa V_+\over3}
   {\rbar}^2 \right) \right] }    \eqno\eq $$
which is exactly the same result as in the previous section.

In the wall region, we expand the integral~\bb\ to first order in
$\alpha$. This gives
$$  B_{\rm wall}=4\pi^2\rbar^3\int d\xi(V_b-V_f)
   +4\pi^2\alpha\rbar^3\int d\xi\bigl[(\sigma^2-8\kappa\sigma V)_{bounce}
   -(false)\bigr]   \eqn\wallhack  $$
At first sight one might think that the first integral is the unperturbed
contribution with $\alpha=0$ and the second integral is the first order
correction. However, the first integral also contains a first order correction
because $V$ is a function of $\phi$ and when $\alpha\ne0$ the dependence of
$\phi$ on $\xi$ changes.
Let us investigate the equations of motion as a power series in
$\alpha$. From Eqs.~\mmotsigma\ and \mmotphi, the equations of motion
in the thin-wall approximation are
$$\eqalign{  \sigma  &  =\kappa(4V+{\phi'}^2)+O(\alpha)  \cr
 \phi'' & ={dV\over d\phi}+4\alpha\kappa\sigma'\phi'-4\alpha\kappa\sigma{dV
  \over d\phi}+O(\alpha^2)
}  \eqn\hackone  $$
Multiplying the second equation by $\phi'$, we get
$$ \Bigl[{1\over2}{\phi'}^2-V\Bigr]'=4\alpha\kappa\Bigl[\sigma'{\phi'}^2-\sigma
V'\Bigr]+O(\alpha^2) \eqn\hacktwo  $$
To first order in $\alpha$, this implies
$$ {1\over2}{\phi'}^2=V-V_0+O(\alpha) \eqn\hackthree $$
Together with the first equation in~\hackone, this leads to
$$ \sigma'{\phi'}^2-\sigma V'=\kappa\bigl(3U^2-4V_0U\bigr)'+O(\alpha)
	\eqn\hackfour  $$
Putting this back in~\hacktwo\ and integrating, we get
$$ {\phi'}^2=2U\bigl(1+12\alpha\kappa^2U-16\alpha\kappa^2V_0\bigr)+O(\alpha^2)
  \eqn\hackfive $$
The first term of~\wallhack\ is
$$ 4\pi^2\rbar^3\int d\xi\bigl(V_b-V_f\bigl)
	=4\pi^2\rbar^3\int{d\phi\over\phi'}\bigl(V_b-V_f\bigr)
  \eqno\eq  $$
Using Eq.~\hackfive\ , we find the first order correction to this is
$$ 4\pi^2\rbar^3\int d\phi\ \alpha\kappa^2\sqrt{2U}(-3U+4V_0) \eqno\eq $$
The contribution to the first order correction from the second integral is
easily found to be
$$ 4\pi^2\rbar^3\alpha\kappa^2\int d\phi\ \sqrt{2U}(-6U-16V_0)  $$
Therefore the first order correction due to the wall region is given by
$$ B_{\alpha|{\rm wall}}= -12\pi^2\alpha\kappa^2\rbar^3\int
  d\phi\ \sqrt{2U}(3U+4V_0)  \eqno\eq$$
Again, this is exactly the same result as in last Section.
Although $\rbar$ is determined in the new metric now, the difference between
the old metric and the new one is first order in $\alpha$. So it does not
invalidate our conclusion.

Therefore, we get exactly the same results by using two completely
different methods. This also proves the validity of the conformal
transformation we discussed in Section II.

I thank Erick Weinberg for suggesting
this subject and for constant discussion and encouragement.

\vfill\eject
\centerline{REFERENCE}\par

\frenchspacing
\item{[1]} S. Coleman, \pr{15}{2929}77.
\item{[2]} C. G. Callan and S. Coleman, \pr{16}{1762}77.
\item{[3]} S. Coleman and F. De Luccia, \pr{21}{3305}80.
\item{[4]} S. Coleman, V. Glaser, and A. Martin, Commun. Math. Phys. {\bf 58},
     211(1978).
\item{[5]} B. Whitt, \pl{145B}{176}84.
\item{[6]} S. Parke, \pl{121B}{313}83.
\item{[7]} Y.~Wang, \pr{42}{2541}90.
\item{[8]} D. La and P. J. Steinhardt, \prl{62}{376}89; D. La and
	P.~J.~Steinhardt, \pl{220B}{375}89.
\item{[9]} P. J. Steinhardt and F. S. Accetta, \prl{64}{2740}90.

\end